\documentclass[epj,english,twocolumn]{webofc}
\usepackage[varg]{txfonts}   % Web of Conferences font
\usepackage[colorlinks=true,citecolor=blue,linkcolor=blue,anchorcolor=blue,filecolor=blue,urlcolor=blue]{hyperref}        % hyperref package
\usepackage{multirow}
\woctitle{Heavy Ion Accelerator Symposium 2019}
\begin{document}
\title{Role of the Surface Energy in Heavy-Ion Collisions}

\author{P. D. Stevenson\inst{1}\fnsep\thanks{\email{p.stevenson@surrey.ac.uk}}}

\institute{Department of Physics, University of Surrey, Guildford, GU2 7XH, UK}

\abstract{%
The surface energy is one of the fundamental properties nuclei, appearing in the simplest form of the semi-empirical mass formula.  The surface enery has an influence on e.g. the shape of a nucleus and its ability to deform.  This in turn could be expected to have an effect in fusion reactions around the Coulomb barrier where dynamical effects such as the formation of a neck is part of the fusion process.   Frozen Hartree-Fock and Time-Dependent Hartree-Fock calculations are made for a series of effective interactions in which the surface energy is systematically varied, using $^{40}$Ca + $^{48}$Ca as a test case.  The dynamical lowering of the barrier is greatest for the largest surface energy, contrary to naive expectations, and we speculate that this may be due to the variation in other nuclear matter properties for these effective interactions
}
\maketitle
\section{Introduction}
\label{intro}
The role of the effective interaction in heavy-ion reactions \cite{Stevenson2019} has been an evolving story over the history of using effective interactions in microscopic reaction theory.  Here we concentrate on those theories built on the basic time-dependent mean-field picture -- Time-Dependent Hartree-Fock (TDHF) \cite{Simenel2018,Sekizawa2019,Nakatsukasa2016,Stevenson2019b}, though the interplay of delvopments to forces and implementation applies elsewhere, too.  The interaction can influence dynamics directly, e.g. through excitation of time-odd fields \cite{Maruhn2006,Stevenson2012}, or because of the importance of the strength of particlar interaction terms \cite{Umar1986,Reinhard1988,Umar1989,Dai2014a,Xu2015,Stevenson2015,Guo2018}.  The influence of the interaction on reaction properties can also arise via the effect on structure properties, such as a variation in symmetry energy leading to changes in neutron radii and a corresponding effect on fusion cross section \cite{Reinhard2016}.  In the spirit of this last study, we make use of a published set of effective interactions in which the surface energy is systematically varied \cite{Jodon2016}, and apply them in heavy-ion fusion reactions around the Coulomb barrier, choosing $^{40}$Ca + $^{48}$Ca reactions as a representative example. 

In section \ref{sec:surface} we give a discussion of the surface energy, and the particular effective interactions used , in section \ref{sec:results} we present a brief description of the methods we use to look at fusion reactions, along with the results obtained of fusion barrier heights, and a discussion of the meaning of these results. Finally some concluding remarks are made in section \ref{sec:conclusions}

\section{Surface energy and Skyrme parameter sets}\label{sec:surface}

\begin{figure}[!tb]
\centering
\includegraphics[width=8cm]{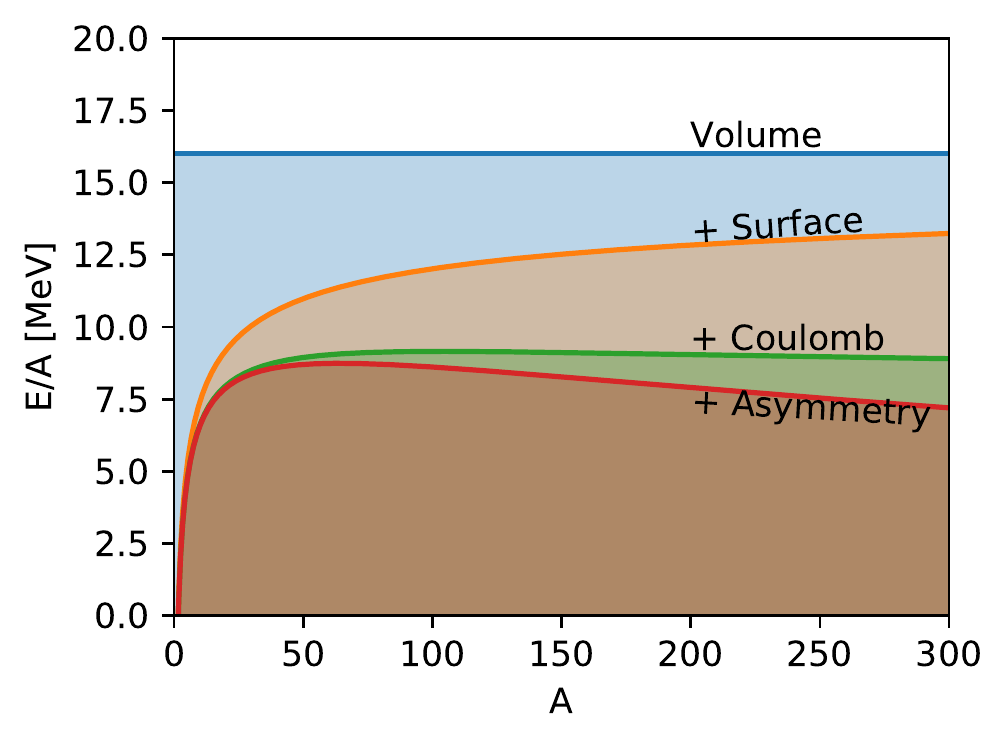}
\caption{Contributions to nuclear binding from different terms in the semi-empirical mass formula.  Starting from the Volume term, each successive lower line cumulatively adds all previous terms, so the line labelled ``+ Asymmetry'' includes all four terms.  The mass formula coeffecients used are as in (\ref{eq:a1}) and (\ref{eq:a2}) except for $a_\mathrm{surf}=-18.5$ MeV.}
\label{fig:semf}       % Give a unique label
\end{figure}

The semi-empirical mass formula (SEMF), in its basic Bethe-von Weizs\"acker form, can be written \cite{Greiner1996}, omitting the pairing term, 
\begin{equation}
  B(A) = a_{\mathrm{vol}}A + a_{\mathrm{surf}}A^{2/3} + a_{\mathrm{coul}}Z^2A^{-1/3} + a_{\mathrm{sym}}\frac{(N-Z)^2}{A},
\end{equation}
with
\begin{align}
  &a_{\mathrm{vol}}\simeq -16\,\mathrm{MeV},\qquad a_\mathrm{surf}\simeq20\,\mathrm{MeV},\label{eq:a1}\\
  &a_{\mathrm{coul}}\simeq 0.751\,\mathrm{MeV},\qquad \mathrm{and}\qquad a_\mathrm{sym}\simeq21.4\,\mathrm{MeV},\label{eq:a2}
  \end{align}
  where $a_\mathrm{vol}$ is the volume term, $a_\mathrm{surf}$ the surface term, $a_\mathrm{coul}$ the Coulomb term, and $a_\mathrm{sym}$ the symmetry term, also known as the asymmetry term.  The values of these terms given in equations (\ref{eq:a1}) and (\ref{eq:a2}) \cite{Greiner1996} are indicattive one obtained from fitting observed binding energies, but one can obtain different values depending on the details of how one performs the fit, by invoking a more detailed SEMF with correction terms, or by relating the various coefficients to other observables besides binding energies \cite{Dutra2012,Danielewicz2003}.  The contribution of the terms, as a function of $A$ is shown in figure \ref{fig:semf} where a smooth function of $Z$ in terms of $A$ is chosen to maximize binding and to give a single line.

The coefficients of the mass formula are intimitely related to the properties of nuclear matter, with the volume term representing the binding per nucleon in infinite nuclear matter, in which each nucleon feels the attraction of its nearest neighbours which surround it.  The surface term, which corrects for those nucleons on the surface which lack some nearest neighbours and hence feel less attraction, is strictly absent in infinite nuclear matter, but can be studied in systems of semi-infinite nuclear matter in which there is a surface, yet nuclear matter extending infinitely in one direction allowing the simplifications that arise in infinite systems, such as washing out complications of shell effects that arise in finite nuclei.  Study of the surface energy acts as a proxy linking nuclear forces, nuclear matter properties, and the properties of real nuclei.

In order to better understand the role of the surface energy in finite nuclei, Jodon \textit{et al}$.$ \cite{Jodon2016} performed fits of the parameters in effective interactions of the Skyrme type \cite{Vautherin1972,Stone2007} in which the surface energy was systematically varied between the fits.  It is not unambiguous how to extract the surface energy from the effective interaction, but for each method used, the fitted forces, labelled SLy5s\textit{n} for $n=1..8$, vary between about 18.0 MeV and 19.4 MeV, in order with $n=1$ havin the smalled surface energy.   We use these parmeter sets in the work presented here.  

\section{Frozen Hartree-Fock and time-dependent Hartree-Fock}\label{sec:results}
In the Time-Dependent Hartree-Fock (TDHF) approximation to a heavy-ion collision, sets of single particle wave functions for each nucleus involved in the collision evolve in time as driven by the moving mean-field and density profile.  As the nuclei involve begin to touch, so the tails of the single particle wave functions overlap and begin the rich dynamics of the reaction process, changing the density profile of the colliding nuclei, forming a neck, and initiating the reaction.

As an initial and simple approach to gauge the dynamics, the Frozen Hartree Fock (FHF) approximation is sometimes used to map out the ion-ion potential without the complexity of the dynamically-evolving single particle states.  In the FHF approximation the ground state sets of wave functions of the colliding nuclei are placed at a fixed set of separations and the resulting energy of the system calculated.  From this is subtracted the energy of the individual isolated nuclei, and one is left with an interaction potential between the two ions.  Formally \cite{Simenel2018},
\begin{equation}
V_{FHF}(\mathbf{R}) = \int d\mathbf{r} \mathcal{H}[\rho_1(\mathbf{r})+\rho_2(\mathbf{r}-\mathbf{R})] - E[\rho_1]-E[\rho_2]
\end{equation}
where the energies are expressed in terms of an energy density functional \cite{Bender2003}
\begin{equation}\label{eq:vfhf}
E[\rho] = \int d\mathbf{r} \mathcal{H}[\rho(\mathbf{r})].
\end{equation}

\begin{figure}[!tb]
\centering
\includegraphics[width=7cm]{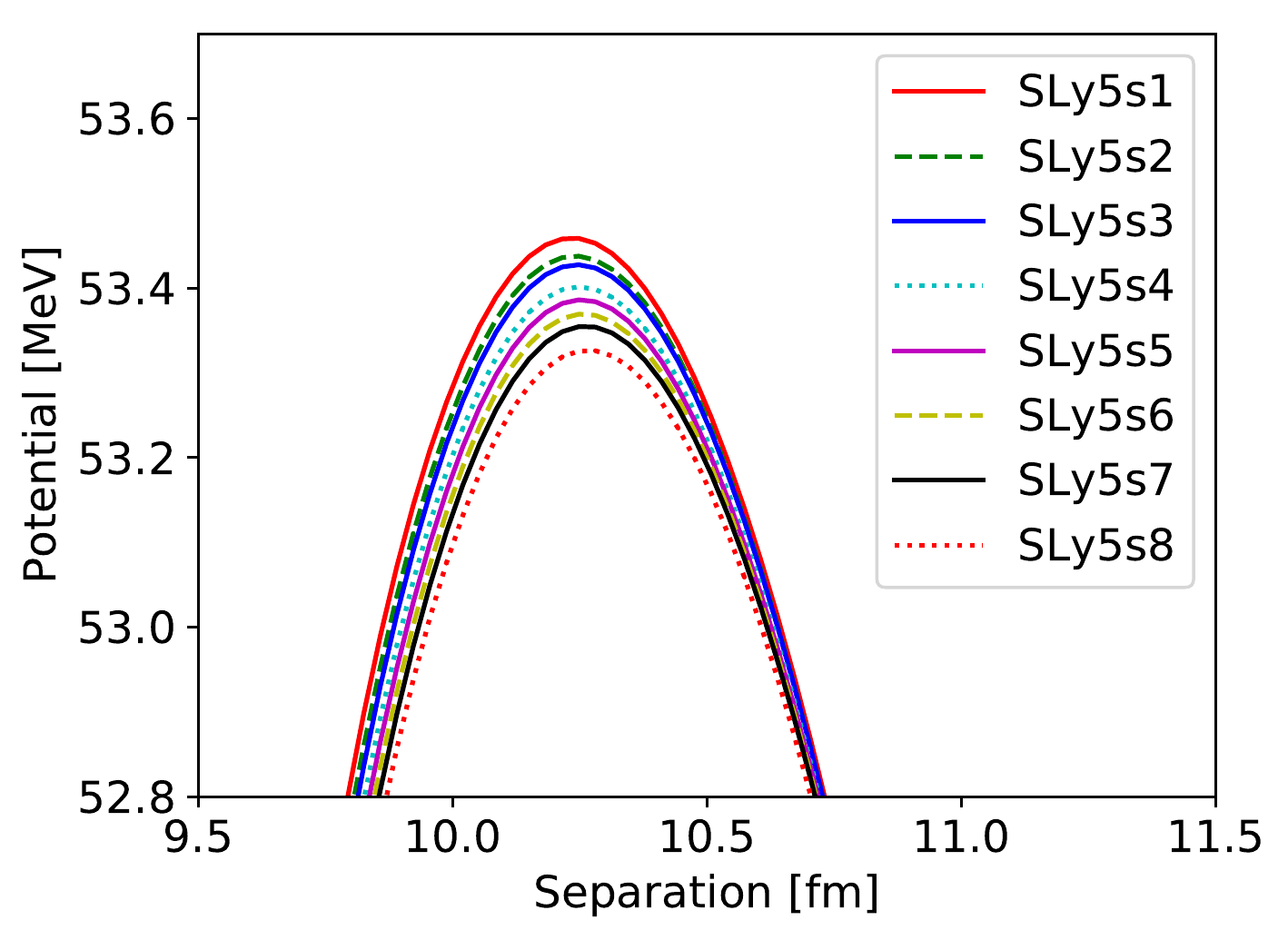}
\caption{The Coulomb barrier, $V_{FHF}$ in (\ref{eq:vfhf}), for $^{40}$Ca + $^{48}$Ca reactions using the Frozen Hartree-Fock (FHF) prescription for the eight SLyS{\textit n} interactions.  The highest barrier is for SLy5s1, and the lowest for SLy5s8, with a monotonic change in barrier height as the forces are traversed in index order.}
\label{fig:barrier}       % Give a unique label
\end{figure}

These FHF calculations give a picture of the Coulomb barrier, and hence the minimum energy needed to overcome it in a fusion reaction.  The can be used as a guide to begin TDHF calculations, as well as a benchmark against which to compare the richer dynamics of TDHF.  For example, one might reasonably expect that when full dynamics is permitted, the ability of the pre-colliding nuclei to change shape will explore a path in which the nuclei can fuse at a lower energy than predicted by FHF.  On the other hand, one may find effects in which early transfer of particles actually hinders fusion compared with the simple FHF picture \cite{Vo-Phuoc2017}.  In any case, the comparison may be physically instructive, and we perform both sets of calucaltions here; FHF and full TDHF calculations to compare barrier heights.

\begin{figure*}[!tb]
\centering
\includegraphics[width=\textwidth]{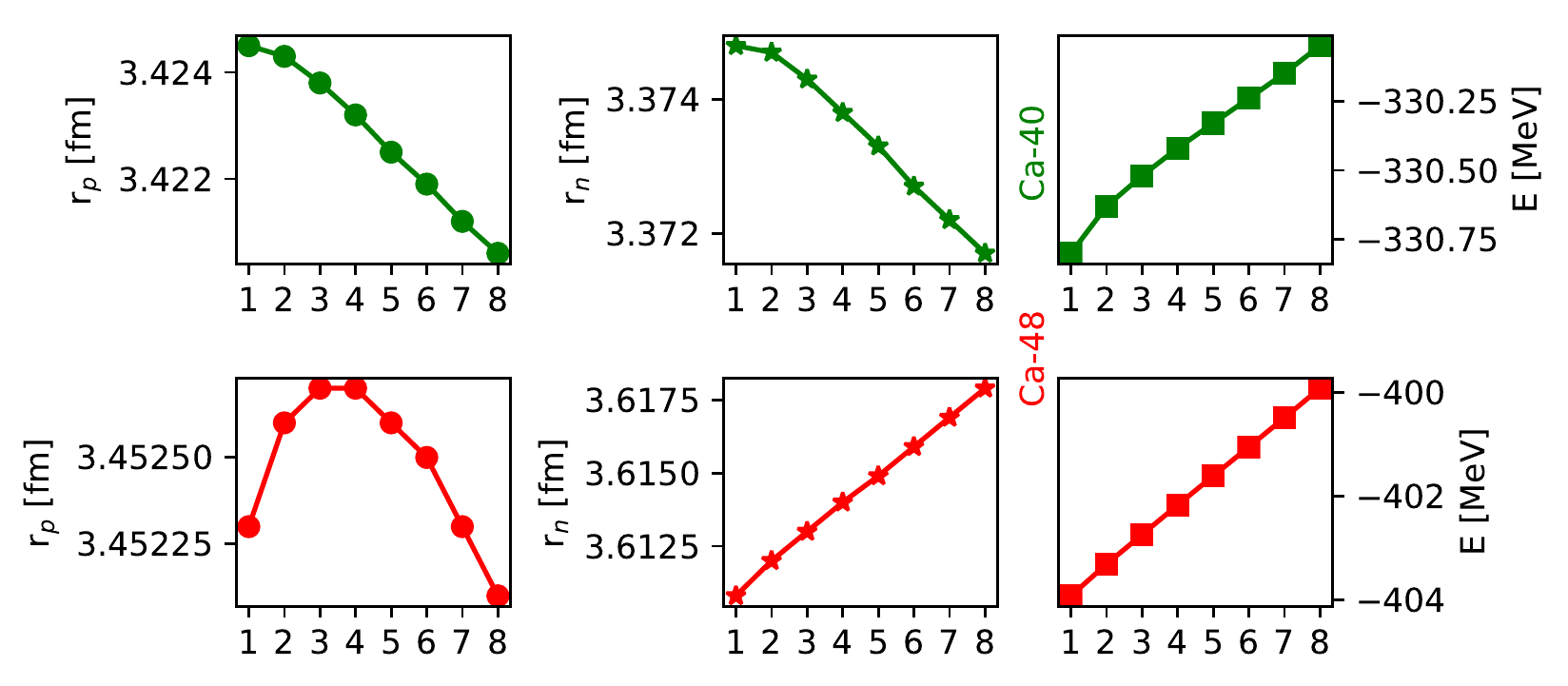}
\caption{Ground state properties for $^{40}$Ca (top row) and $^{48}$Ca (bottom row) using the different SLy5s$n$ forces for $n=1..8$.  The observables are proton radius (left column), neutron radius (middle column) and energy (right column).}
\label{fig:groundstates}       % Give a unique label
\end{figure*}

Figure \ref{fig:barrier} shows the Frozen Hartree-Fock barriers in $^{40}$Ca + $^{48}$Ca.  One sees that the smallest surface energy gives rise to the largest barrier.  This does not include any effects of shape polarisation, in which one might expect the lower surface energy to decrease the barrier.

To help understand the effect of the different interactions on the FHF results, the ground state properties are visualised in figure \ref{fig:groundstates}.

One sees here that depending on the nucleus, the systematic change in surface energy corresonds to specific changes in these observables, but that the results can differ for different nuclei.  Incresing the surface energy decreases the neutron radius for $^{40}$Ca while increasing it for $^{48}$Ca, for example.  The complicated interplay here of density distributions and extents will affect the barrier as nuclei are placed close together in the FHF procedure.  The energies, too, will affect the barrier heights. 

\begin{table}[!tb]
\centering
  \caption{Barrier heights for $^{40}$Ca + $^{48}$Ca with FHF and TDHF methods for Skyrme forces SLy5s$n$ for $n=1..8$.}
  \begin{tabular}{ccc}
  \hline
    Force & \multicolumn{2}{r}{Barrier Heights [MeV]} \\
    \hline
    & FHF & TDHF \\
SLy5s1 &53.457 &51.885$\pm$0.005 \\
SLy5s2 &53.437 &51.855$\pm$0.005\\
SLy5s3 &53.427 &51.825$\pm$0.005\\
SLy5s4 &53.401 &51.795$\pm$0.005\\
SLy5s5 &53.386 &51.765$\pm$0.005\\
SLy5s6 &53.369 &51.735$\pm$0.005\\
SLy5s7 &53.355 &51.705$\pm$0.005\\
SLy5s8 &53.326 &51.665$\pm$0.005\\
\hline
\end{tabular}
  \label{tab:barriers}
\end{table}

\begin{figure}[!tb]
\includegraphics[width=0.45\textwidth]{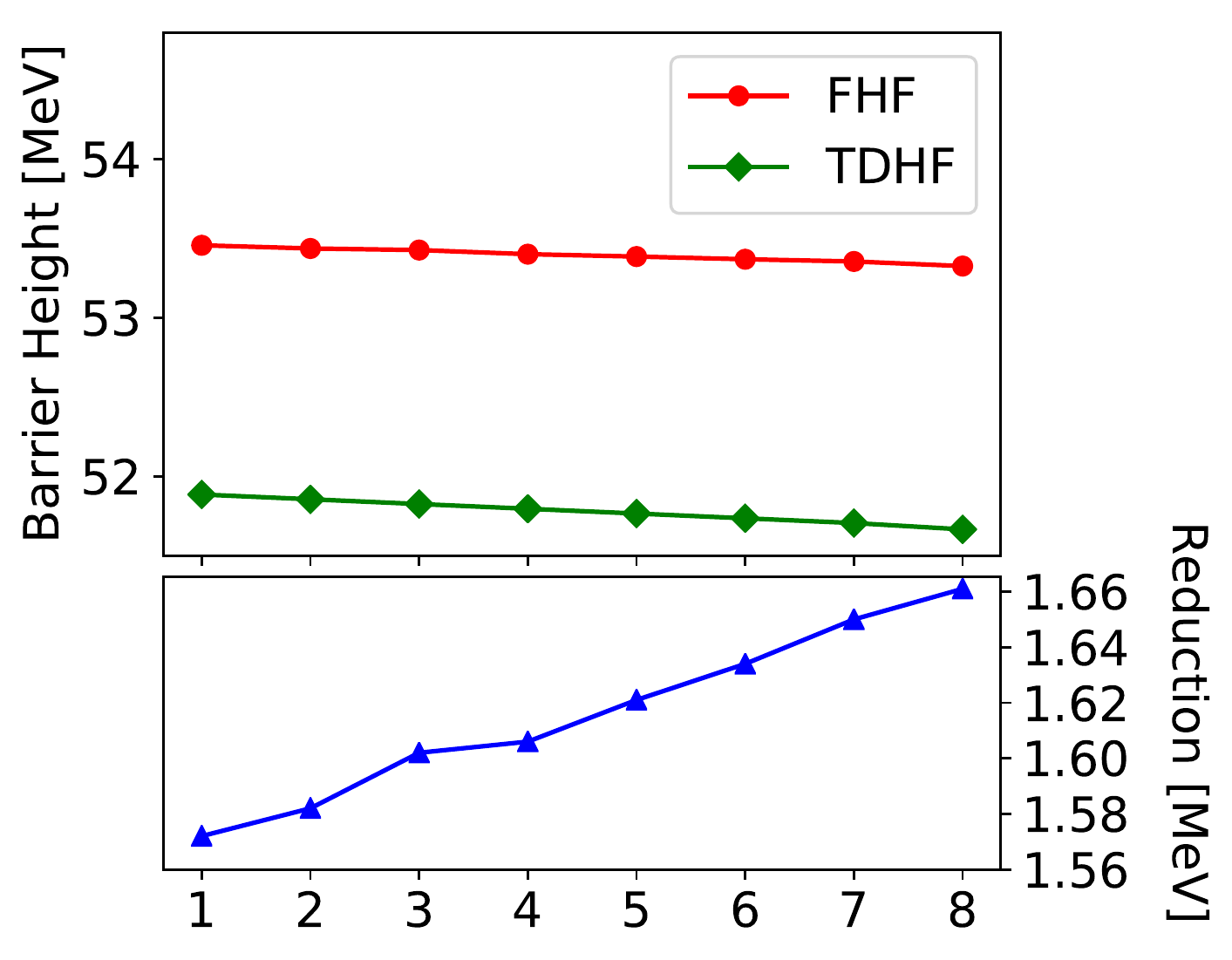}
  \caption{The barrier heights as given by FHF and TDHF (top panel) and the difference between them (lower panel) for $^{40}$Ca + $^{48}$Ca fusion reactions.}
  \label{fig:barriers}
\end{figure}

To understand if the surface energy has a decisive dynamical effect in the barrier heights, a comparison is made between the FHF calcualtion and a TDHF calculation.  In the latter, the nuclei may (indeed, do) deform as they approach each other.  The extent to which a neck forms to aid fusion could be expected to depend on how easiliy the nucleus forms a surface, or what the ratio of the surface to volume energy is.

We therefore find the barrier height in a TDHF calculation by finding the lowest energy at which a fusion reaction proceeds for each of the eight forces in the $^{40}$Ca + $^{48}$Ca reaction.  These are tabulated in Table \ref{tab:barriers} and compared with the FHF results in Figure \ref{fig:barriers}.  As expected, the dynamical effects allowed in TDHF reduce the barrier height.

As seen from the lower panel of Figure \ref{fig:barriers}, the dynamical effect \textit{increases} as the surface energy increases - i.e. dynamical effects cause a greater lowering of the barrier for higher surface energy.  On the face of it, these results would seem to run counter-intuitively to the idea that a higher surface energy gives a stiffer equation of state for the nuclear matter in the nucleus against forming a surface, or a neck in a fusion reaction.

One explanation of this effect could be that the asymmetry energy is not constant across the SLy5s$n$ force series.  Indeed, the asymmetry energy increases by about 1 MeV from SLy5s1 to SLy5s8.  This may explain the increasing trend of $r_n-r_p$ for $^{48}$Ca as $a_\textrm{surf}$ increases, and this in itself may account for a greater dynamical effect \cite{Klupfel2009}.  

We are making further explorations of the effects of the surface energy in dynamics, including in heavier nuclei, whose analysis will appear in a follow-up publication to this conference proceeding.

\section{Conclusion}
\label{sec:conclusions}

We have performed Frozen Hartree-Fock and Time-Dependent Hartree-Fock calculations for fusion of $^{40}$Ca + $^{48}$Ca using a series of Skyrme force parameterisations in which the surface energy has been systematically adjusted.  We find a monotonic increase in the magnitude of the dynamical reduction in the barrier height between FHF and TDHF as a function of increaing surface energy.  This is contrary to the naive expectation that a higher surface energy will make it harder to form a neck and so will not lower the barrier by such a large amount.  We speculate that this could be because the set of Skyrme forces also feature a variation of other nuclear matter properties (such as the symmetry energy) and a more complete anaylsis is therefore needed.

It is clear from the original paper in which these forces are introduced \cite{Jodon2016} that some properties at very high deformation (e.g. potential energy surface in $^{240}$Pu are significantly affected by the surface energy and we plan further studies, such as to look at giant resonances and fission lifetimes in heavy nuclei.

% BibTeX or Biber users please use (the style is already called in the class, ensure that the "woc.bst" style is in your local directory)
%\bibliography{../../pds-bib.bib}

\end{document}